# Statistical Sensitiveness for Science

Jose D. Perezgonzalez

Massey University, New Zealand

**Author's note:**
Jose D. Perezgonzalez, Business School, Massey University.
Correspondence concerning this article should be addressed to Dr. Jose D. Perezgonzalez, Massey University, Social Sciences T8.18, Manawatu Campus, P.O. Box 11-222, Palmerston North 4442, New Zealand. Phone: +6463505326, E-mail: j.d.perezgonzalez@massey.ac.nz

# Abstract

Research often necessitates of samples, yet obtaining large enough samples is not always possible. When it is, the researcher may use one of two methods for deciding upon the required sample size: rules-of-thumb, quick yet uncertain, and estimations for power, mathematically precise yet with the potential to overestimate or underestimate sample sizes when effect sizes are unknown. Misestimated sample sizes have negative repercussions in the form of increased costs, abandoned projects or abandoned publication of non-significant results. Here I describe a procedure for estimating sample sizes adequate for the testing approach which is most common in the behavioural, social, and biomedical sciences, that of Fisher's tests of significance. The procedure focuses on a desired minimum effect size for the research at hand and finds the minimum sample size required for capturing such effect size as a statistically significant result. In a similar fashion than power analyses, sensitiveness analyses can also be extended to finding the minimum effect for a given sample size a priori as well as to calculating sensitiveness a posteriori. The article provides a full tutorial for carrying out a sensitiveness analysis, as well as empirical support via simulation.

# Introduction

Imagine for a moment that you want to test the effect of a novel treatment on a dependent variable of interest. You intend to carry out an experiment with two independent groups using a test of significance. How big ought your sample to be?

In order to determine the appropriate sample size, you may consider what is typical in your discipline, perhaps even cost or opportunity (a procedure which is still quite common, as seen in recent articles published in *Behaviour Research Methods*, such as Caspar et al., 2015, Roncero & Almeida, 2015, and Widmer et al., 2015). Let's say you opt for a sample of 30 participants. Reassured by tradition or good budgeting, you brace yourself for the best and hope for a statistically significant result. You may, however, feel that this is not a very scientific way of setting a sample size and wish for a more reliable procedure.

Sample size estimation for power offers that reliable procedure for setting sample sizes, and there now exist several desktop and online programs for effortless estimation. You thus download and run G*Power (see also Faul, Erdfelder, Lang, & Buchner, 2007) and pretty much you get stuck: You can certainly decide the appropriate test, whether to use a one-tailed or a two-tailed test, and what information to enter for alpha, allocation ratio, and, even, power; however, you have absolutely no idea about which effect size to use—how could you possibly know given that this is a novel treatment! It does not really matter how hard you think about it and how reasonable your arguments may be, at the end you are left to just one resource: Decide on an effect and see what sample size the program returns; if too large, you may fiddle with the parameters until you reach a satisfactory compromise (something recommended by, for example, Cohen, 1988, and Lenth, 2001). In hindsight, you probably still feel that this is neither more scientific nor more reliable than the rules-of-thumb used earlier, yet it typically commands larger samples.

In both cases get your estimate right, and you are on your way to a successful publication. Get it wrong and use more participants than needed, and you would have brought about unnecessary harm (e.g., in the form of higher costs, animals unnecessarily sacrificed, or excess patients unnecessarily harmed). Even worse, get it really wrong by using less participants than needed, and the harm done spills to the whole sample as the research project is filed away because results may not be publishable, unless you manage to motivate yourself to write up your "negative" results as well as manage to get your paper into one of the few journals that publish nonsignificant results.

Is there a way out of this conundrum? Yes, there is; therefore the goals of this article: To describe the bases of an alternative procedure for estimating sample sizes when we lack enough knowledge about population effect sizes as for using power estimation, to support such procedure with empirical results from a series of simulations, and, finally, to provide a tutorial on how to reliably estimate sample sizes using such alternative procedure.

As introduced above, accurate estimation of the required sample size for a research project may mean the difference between a significant test result and one confined to the file drawer, especially in the behavioural (Ioannidis, Munafò, Fusar-Poli, Nosek, & David, 2014), social (Franco, Malhotra, & Simonovits, 2014), and biomedical sciences (Papageorgiou, Papadopoulos, & Athanasiou, 2014). Two common approaches to sample estimation are rules-of-thumb and estimations for power.

Rules-of-thumb represent quick heuristics for setting sample sizes. They may be borne out of statistical principles (e.g. a sample size of 30 is about the minimum needed for rendering normal sampling distributions of means independently of whether the sample itself is normal; Aron, Coups, & Aron, 2013; also Crawley, 2014). Or they may be linked to particular tests (e.g., to $\chi^2$; Van Voorhis & Morgan, 2007), to particular procedures (e.g., to a factor analysis; Tabachnick & Fidell, 2001), or to particular goals (e.g., to costs; Van Belle, 2008). Often enough, they are simply handed down as such rules-of-thumb (e.g., Cross Validated, 2014, "Rules of Thumb for Minimum Sample Size", or Tools4Dev, 2014, "How to Choose a Sample Size"). These rules are independent of any particular research context and, thus, represent an imprecise approach to setting sample sizes.

Sample size estimations for power were popularized by Cohen (1988) with his book *Statistical Power Analysis for the Behavioral Sciences*, which provided a one-stop shop for also calculating the power of tests a posteriori and the size of effects—the book is nowadays supplanted by more convenient tools like proprietary computer software, such as G*Power and SPSS SamplePower, and online calculators, such as those provided on DanielSopper.com. Unlike rules-of-thumb, power-based estimations are tailored to particular research projects. Indeed, power analysis provides such good value to behavioural, social and biomedical research that even grantees may ask for power analyses to be included with grant applications (e.g., National Institutes of Health, 2015; Voelker, 2014; Yale Institution for Social and Policy Studies, 2015).

Power, however, is circumscribed to Neyman and Pearson's approach to data testing (1933) and, thus, is mostly relevant when researchers work with competing precise hypotheses, know the effect size in the population, assume a repeated sampling framework, and are concerned with controlling Type I and Type II errors in the long run (Perezgonzalez, 2015a). Actually, Neyman-Pearson's approach offers such good a priori control of the research environment that it is ideal for designing simulations (e.g., Jo, 2002; Rast & Hofer, 2014; Siemer & Joormann, 2003; Simonsohn, Nelson, & Simmons, 2014; Stanley & Spence, 2014), even when sometimes such simulations may be used for testing unsuitable research scenarios (e.g., Perezgonzalez, 2015b, c).

As it happens, most research in the behavioural, social and biomedical sciences does not work with repeated sampling from the same population—often enough, it does not even define the effect size of interest for differentiating between hypotheses nor control Type II errors a priori. Thus, it cannot be said to follow Neyman-Pearson's approach. Instead, most of it defaults to Fisher's approach (e.g., 1960): Testing research data against a single null hypothesis and ascertaining the ad hoc probability of the research results against such hypothesis[1].

When using Fisher's tests of significance, power analyses are philosophically meaningless. Such analyses may also force researchers to guess population effect sizes and may turn costly either because of an overestimation of the sample size so calculated—when the effect size is underestimated—or because of underpowered research—when the effect size is overestimated. Furthermore, estimating effect sizes when no knowledge about the population exists, whether based on armchair thinking, guesses, hopes or wishful thinking, is also foreign to a conscientious researcher and ethically questionable, which partly explains why power estimations have not yet managed to supplant rules-of-thumb in current research sampling (e.g., Button et al., 2013; Maxwell, Lau, & Howard, 2015; Vadillo, Konstantinidis, & Shanks, 2015). Indeed, ignoring power does not necessarily affect significance, as a test can return a significant result even if its power is low.

---

[1] Fisher's and Neyman-Pearson's approaches are typically mixed up into the philosophically incoherent null hypothesis significance testing (NHST) procedure (Gigerenzer, 2004; Perezgonzalez, 2015a). In this article I will avoid references to NHST, referring to the sounder approaches of Fisher and of Neyman-Pearson, instead.

What I discuss in this article is that sensitiveness, not power, is the appropriate construct to attend to in the case of Fisher's approach:

> By increasing the size of the experiment [either by enlargement or repetition], we can render it more sensitive, meaning by this that it will allow of the detection . . . of a quantitatively smaller departure from the null hypothesis . . . We may say that the value of the experiment is increased whenever it permits the null hypothesis to be more readily disproved. (Fisher, 1960 pp. 21-22)

Unfortunately, Fisher did not provide a method for controlling sensitiveness, thereby the purpose of this tutorial.

## The relationship between power and sensitiveness

Neyman, in particular, found Fisher's dismissal of power troubling:

> It is now appropriate for me to mention a problem which obviously belongs to the category of experimental tactics, but is missing in Fisher's writings . . . The problem is that of the power of the tests contemplated for the treatment of the given experiment. The consideration of power is occasionally implicit in Fisher's writings, but I would have liked to see it treated explicitly. (1967, p. 1459)

Despite Neyman's perception of the similarity between power and sensitiveness, Fisher's dismissal of power is understandable given that sensitiveness is rooted in the different philosophical position and research goal of tests of significance.

As said earlier, power is circumscribed to Neyman-Pearson's approach to data testing (1933) and, thus, is most relevant when research has the following parameters: A known effect size in the population which differentiates between two statistical hypotheses (this effect size is provided by the alternative hypothesis); a Type I error probability (this is the probability of wrongly rejecting the main hypothesis under test in the long-run, under repeated sampling from the same population); and a Type II error probability (this is the probability of wrongly rejecting the alternative hypothesis in the long-run, under repeated sampling from the same population). The purpose of testing data is, primarily, to decide which hypothesis explains the research data better—a goal which Fisher considered more appropriate for industrial quality control than for contemporary research endeavours (Fisher, 1955).

Fisher's tests of significance, on the other hand, are tools for learning from the data at hand, ad hoc, without any frequentist assumption of repeated sampling from the same population. It works despite not knowing the population effect size, not knowing the parameters of an alternative hypothesis, and not been interested in long-run errors linked to such hypothesis. Furthermore, the only purpose of testing data gets reduced to rejecting an otherwise uninformative null hypothesis (Perezgonzalez, 2015a).

Despite their differences, Neyman-Pearson's procedure and Fisher's procedure are intimately related, inasmuch Neyman-Pearson's necessitates of Fisher's to test the data. In a nutshell, a Neyman-Pearson's test is a transduction of an acceptance philosophy onto a test of significance.

Figure 1 renders Fisher's and Neyman-Pearson's approaches to data testing. Figure 1A represents Fisher's approach: a one-tailed test in a *t* distribution with 162 degrees of freedom. The distribution concerns a null hypothesis ($H_0$) with a conventional area of significance at the 1% level. The level of significance (*sig*) represents a cut-off point in such distribution, thus being also associated to a particular critical value (*CV*), even if this value is largely ignored when interpreting results because it is redundant with the more informative level of significance.

Neyman-Pearson's approach (Figure 1B) is typically represented by the distributions of both a main hypothesis ($H_M$) centered on 'zero' and an alternative hypothesis centered on a different value. The difference in location between these distributions (a.k.a., their degree of overlapping and nonoverlapping, Cohen, 1988) represents the effect size in the population (e.g., Cohen's $d = 0.5$). The test, however, is carried out on the main hypothesis only, which is what Figure 1B represents (the reader may project the distribution of the alternative hypothesis by continuing the dashed lines off towards the right: The distance between the means of the hypotheses is the effect size). The test represented in Figure 1B assumes a 1% Type I error ($\alpha$) and a 20% Type II error ($\beta$).

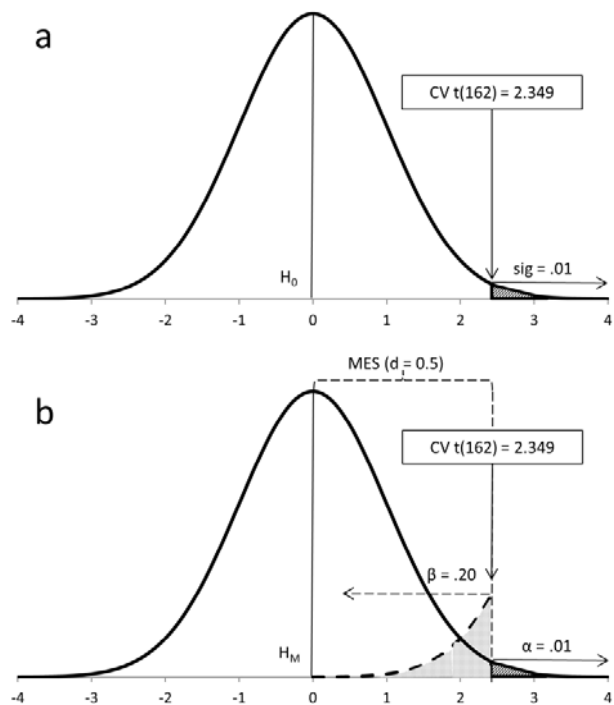

*Figure 1*. Fisher's (a) and Neyman-Pearson's (b) approaches to data testing (*t*-test for independent samples).

As Figure 1 shows, both approaches are technically alike, more so when they coincide in using the same test and distribution. Indeed, Neyman-Pearson's alpha level doubles as the level of significance for the test and cuts the distribution at the same critical value ($CV = 2.349$)[2]. The only difference is that Neyman-Pearson's approach also incorporates relevant information provided by the alternative hypothesis: that is, the probability of making a Type II error in the long run ($\beta$) and the population effect size.

Before continuing our argument, however, a small detour is needed to explain what the effect size is, as this parameter is critical for the abovementioned argument. Let's start with Fisher's tests of significance.

A test of significance is carried out in order to determine whether a phenomenon exists in the population or not. The hypothesis under test is that the phenomenon does not exist (e.g., $H_0 = 0$). However, the hypothesis is tested against a particular frequency distribution, effectively allowing other values than 'zero' to enter the test (i.e., the assumption is that members of a population randomly show differing effects even when, in the aggregate, the phenomenon is not present in such population). Different results within this frequency distribution will have different probabilities of occurring, extreme results being rarer and, thus, having smaller probabilities. The level of significance (e.g., $sig \approx .01$) serves as a convenient cut-off for making our mind in this respect, by comparing the observed probability of our research data ($p$) against the level of significance. A larger $p$-value informs nothing, while the smaller the $p$-value (e.g., $p = .001$) the more confident we can be that the research data do not fit the null hypothesis well.

The distribution shown in Figure 1A represents a $t$ distribution with 162 degrees-of-freedom: It represents the density of frequencies around the mean value of 'zero' and also provides the theoretical location of all possible results ($t$-values become larger the further away from 'zero') and their associated probability ($p$-values become smaller the further away from 'zero'). In a nutshell, each $t$-value has a $p$-value and either can be used instead of the other (the $p$-value being more convenient).

Now, although Fisher did not work with effects sizes, it is possible to derive them from a test (e.g., Cohen, 1988, provided the appropriate formulas). This means that any $t$-value in our distribution can also be associated with a particular effect size in the population—"To this point," said Cohen (1988, p 11) "the ES [can be] considered quite abstractly as a parameter which can take on varying values (including zero in the case of the null case)." It also means that the level of significance is a cut-off in the distribution of $t$'s (and, by extension, of associated $p$'s and effect sizes) in order to determine which shall be taken as evidence against $H_0$ and which shall not. It also implies that $t$'s and $p$'s can be used for estimating the true but unknown effect size in the population, especially when combined into a meta-analysis (Rosenthal, 1984; Braver, Thoemmes, & Rosenthal, 2014).

A similar logic is applicable to Neyman-Pearson's tests (Figure 1B), with the particularity that these tests actually consider two overlapping distributions ($H_M$ and $H_A$), separated by a known effect size. This means that each possible location under $H_M$ will also appear under $H_A$ (albeit with a different $t$-value and $p$-value), and that each effect size under $H_M$ is actually an effect size under the now explicit $H_A$ (because the expected effect under $H_M = 0$). Neyman-Pearson's procedure is nonetheless simplified by only testing the main hypothesis using Fisher's tests of significance. Therefore, the cut-off represented by alpha ($\alpha$ = .01) not only determines which *t*'s (and *p*'s) fall under $H_M$ and which fall under $H_A$, it effectively partitions the effects expected under the alternative hypothesis into those important ones which will be accepted under $H_A$ and those unimportant ones which will be assumed under $H_M$. (In Figure 1B, the information about the population effect size provided by $H_A$ is thus represented by MES, the unimportant portion of the effect size parameter that falls under $H_M$.)

---

[2] Neyman-Pearson's alpha and Fisher's level of significance are often confused and indistinctively called after each other (e.g., Hubbard, 2004). In this article, I will distinctly differentiate between them to avoid confusion (as recommended in Perezgonzalez, 2014).

In finishing our detour and returning to our argument, we can thus compare Figures 1A and 1B and readily ascertain that, apart from the nuances of *β* and *MES* incorporated by Neyman-Pearson's tests, the two approaches are technically alike. Indeed, it is precisely this overall technical similarity that allows resolving the problem of sensitiveness.

Firstly, both Fisher and Neyman-Pearson end up carrying out a (Fisher's) test of significance on a statistical hypothesis centered on 'zero' (Fisher's null hypothesis—$H_0$—and Neyman-Pearson's main hypothesis—$H_M$—, respectively). As shown in Figure 1, for the case when both approaches coincide in using the same test and sample size, the theoretical distribution for both tests is the same.

Secondly, the level of significance (*sig*) and the probability of Type I error (*α*) are technically similar for any individual test: Neyman-Pearson's alpha simply doubles as Fisher's level of significance. As shown in Figure 1, the *sig* / *α* level also results in the same critical value (*CV*) and rejection region for both approaches when using the same distribution.

Thirdly, as only the main hypothesis ($H_M$) is tested under Neyman-Pearson's approach, it makes sense to represent just the proportional effect size, or minimum effect size (*MES*), that falls under $H_M$. As shown in Figure 1B, *MES* shares boundaries with *α, β,* and *CV*, and has the following interpretation: It represents a region of effects sizes under the alternative hypothesis which will be assumed to pertain to the main hypothesis every time the latter is accepted.

Indeed, because *MES* coincides with *CV*, we can see that it is *MES*, not the population effect size, which is used for the test. However, *MES* does not have an inherent meaning under Neyman-Pearson's approach: In our example, it is either $ES = d = 0$ ($H_M$) or $ES = d = 0.5$ ($H_A$) which explains the research results better, not any particular *MES*. Furthermore, altering *α, β, N* or the test (e.g., by using a two-tailed test) will alter *CV* and, consequently, *MES*, yet the population effect size remains the same. Thus, *MES* underlies Neyman-Pearson's tests yet does not need to be derived mathematically because the population effect size is a more informative and convenient statistic. (The interested reader may calculate it as shown later. In the case of Figure 1B, the population effect size of Cohen's $d = 0.5$ renders an *MES* of $d = 0.37$.)

As discussed by Cohen (1988), effect sizes are properties of the populations of reference, not of statistical tests. Therefore, despite the lack of an explicit alternative hypothesis, effect sizes are also coherent constructs under Fisher's approach. A minimum effect size can, thus, be drawn in a Fisher's test distribution, sharing boundaries with *sig* and *CV*.

Unlike Neyman-Pearson's *MES*, however, Fisher's *MES* is not a mathematical boundary between the distributions of two known hypotheses. Instead, it is an *MES* which the researcher decides upon as relevant to her research, independently of the actual distribution of the unknown alternative hypothesis. This *MES* does not depend on previous knowledge nor requires thinking about a population of reference: It may as well be decided upon ad hoc and it certainly accommodates wishful thinking as easily as more considered reasoning. This is so because what is decided upon is the minimum effect size that would justify the cost of the research and its future implementation. (This idea is similar to Lenth's, 2001, lower bound on the effect size, for which he even proposed a procedure to help determine it, albeit in the context of power estimation.) In a nutshell, the researcher is not simply thinking about obtaining any statistically significant result—a measure of direction only—but about obtaining a statistically significant result for a desired *MES*. Said otherwise, the researcher is not simply aiming to differentiate between vague noteworthy results and those she may ignore, but to decide how much is noteworthy—her desired minimum effect size and larger effects—and to ensure that she has a large enough sample to capture such effects as statistically significant results.

Therefore, knowing the *CV* that corresponds to the desired significance level allows for *MES* to be derived mathematically—via the corresponding formulas for post hoc estimation, see Supplemental Table 1—which, in our case, is equal to Cohen's $d = 0.37$. (Notice that post hoc estimation of population effect sizes makes sense under Fisher's, as the approach works with the assumption that nothing is known about such effect sizes. This is unlike Neyman-Pearson's approach, which starts with known effect sizes and, consequently, post hoc calculations of the same are meaningless.)

Fourthly, and finally, the problem of *β* cannot be resolved as it depends on an alternative hypothesis that Fisher's approach does not consider. However, this problem can be circumvented with a procedure that estimates sample size based on just *sig* and *MES*. This procedure estimates sensitiveness, defined as the minimum sample size required for returning a significant test result for a given (minimum) effect size.

## Simulations using sensitiveness, power, and rules-of-thumb

In wishing to place the procedure for estimating sensitiveness at the end of the article, where it will be much easier to locate later on, I put the proverbial "cart before the horse" and present the results of several simulations contrasting sample size estimations for sensitiveness, for power, and when using a rule-of-thumb.

Traditional simulations based on repeated sampling from the same population are suitable for Neyman-Pearson's approach—in which population effect sizes are known—but not so for Fisher's—in which population effect sizes are not known. Therefore, I designed a more suitable simulation program, based on generating multiple populations with relatively uncertain parameters from which to extract the samples. The overall simulation program was run 'manually' and comprised eight simulations (called 'studies'), 43 independent populations per study (thus, yielding 344 independent populations overall), three conditions per population (sensitiveness, power, and rule-of-thumb), and one independent sample per condition (thus, yielding 129 samples per study and 1,032 samples in total—see Supplemental Material 3 for further information).

Each research condition comprised a *t*-test for independent groups using two groups of equal size: the rule-of-thumb condition used a total sample size of 30 (i.e., $n = 15$ per group), the power condition used a total sample size of 102 (i.e., $n = 51$ per group), and the sensitiveness condition used a total sample size of 48 (i.e., $n = 24$ per group). The power condition represented the sample size recommended for capturing Cohen's $d = 0.5$ with 80% power, and the sensitiveness condition represented the sample size recommended for capturing Cohen's $d = 0.5$ as a statistically significant minimum effect size (*MES*).

The data of interest for analysis were the frequencies, or counts, of effect sizes greater than Cohen's $d = 0.495$ (to allow for rounding) which turned out to also be statistically significant (*sig* $\leq .05$). The interest was for the effect size to take precedence over mere statistical significance, the reasoning being that a researcher who does not know the effect size in the population but uses power analyses for determining sample size, most probably decides on an effect size as the minimum size he wishes to find—this effectively meant that both significant yet smaller effect sizes and nonsignificant larger effect sizes were ignored.

For the analysis of frequencies, the target effect size was a medium effect (Cohen's $w = .3$). This *MES* meant to be of enough size that a reasonable researcher would be satisfied that, compared to the smaller sample size recommended by the sensitiveness condition, whatever costs associated with the larger sample recommended by the power condition (slightly over twice as large) would be justified by the benefits it may bring (i.e., 30% or more tests captured). Such *MES* recommended a sample size of 43 per group for pair comparisons ($\chi^2$) which is the sample size used per condition in each of the eight studies.

Frequency analyses ($\chi^2$ for goodness-of-fit) were thus carried out between pairs of conditions using the number of effect sizes greater than $d = 0.495$ that turned up as statistically significant. Collated results for the overall simulation are presented in Table 1 (and for each of the eight studies in Supplemental Table 4).

Table 1

*Significant tests which captured MES d > 0.495 (Frequencies, Percentages, Effect Sizes, and Tests of Significance)*

| Condition | $\Sigma f$ | % | Test | w | $\chi^2$ | p |
|---|---|---|---|---|---|---|
| PWR | 158 | 36.8 | Pwr-Sns | .03 | 0.26 | .61 |
| SNS | 149 | 34.7 | Sns-Thmb | .10 | 2.69 | .10 |
| THMB | 122 | 28.4 | Thmb-Pwr | .13 | 4.63 | .03 |

*Note.* $\Sigma f$ = accumulated frequencies for significant tests capturing d > 0.495; PWR = power-based sampling, SNS = sensitiveness-based sampling, THMB = rule-of-thumb-based sampling.

The simulation results show that, while a larger sample tends to also be more capable of capturing the desired effect sizes as statistically significant, the difference between power-based sampling and sensitiveness-based sampling was rather unimportant ($w = .03$, range [.03, .24], $\chi^2 = .26$, $p = .61$).

Surprisingly enough, although the N = 30-rule-of-thumb-based sampling performed worse than sensitiveness-based sampling ($w = .10$, range [.03, .33], $\chi^2 = 2.69$, $p = .10$) and power-based sampling ($w = .13$, range [.06, .26], $\chi^2 = 4.63$, $p = .03$), the difference in performance was definitively smaller than the target *MES* set for the simulation. However, given that rules-of-thumb sampling recommendations are independent of research context, this type of sampling may prove less reliable than the other two and I shall not discuss it further.

As for the sampling procedures of interest, it is reasonable to conclude that power-based sampling is not sensibly better than sensitiveness-based sampling when used as intended in this article, yet it commands a larger sample size and, thus, is costlier. The results of this simulation support the use of sensitiveness analysis rather than power analysis when research designs are typical of the former. What now follows is a basic tutorial for carrying out sensitiveness analyses.

## Statistical sensitiveness analysis (a tutorial)

**Sample size estimation for sensitiveness**

In order to estimate sensitiveness, I devised the following two-step process, to be iterated until the required sample size for sensitiveness is found:

- The first step is to calculate the critical value of a test that allows for rejecting a research result at a certain level of significance. As the level of significance remains fixed, the only parameter to manipulate is the sample size. This step can be conveniently performed with online calculators (e.g. DanielSoper.com) or by consulting appropriate tables.
- The second step is to calculate the effect size associated with above critical value by using the appropriate formulas (Supplemental Table 1). This step can be conveniently performed with online calculators (e.g. UCCS.edu) or it can be easily automatized using Excel.

For example, let's assume we are interested in estimating the minimum sample size capable of capturing an intergroup difference equal to Cohen's $d = 0.37$ (i.e., a small-to-medium effect size) with a one-tailed *t*-test at the 1% level of significance. We can open DanielSoper's *t* value calculator (http://www.danielsoper.com/statcalc3/calc.aspx?id=10), enter a probability level of .01, then enter any initial value as degrees of freedom (let's say, 100). This will return the corresponding *t* value of 2.36. We then proceed to calculate the associated effect size by using the corresponding formula—see Supplemental Table 1—, which returns a Cohen's $d = 0.47$. This *d* is larger than target *MES*, suggesting that we need a larger sample in order to capture the smaller target. Let's enter $df = 200$ and repeat above steps, which yields $d = 0.33$, smaller than target *MES*. With some fine-tuning, we will soon converge on $df = 162$, thus estimating $N = 164$ ($df = 162 + 2$) as the minimum sample size needed for capturing $MES = d = 0.37$ as a statistically significant result (which is what Figure 1A represents).

Table 2 provides a breakdown of the sample sizes required for sensitiveness and power for Cohen's conventional effect sizes and for several statistical tests (see also Supplemental Table 2). As shown, the required sample sizes for sensitiveness are markedly smaller than those required for power. The reason for these differences is that power targets an effect size in the population—yet obscuring the fact that it tests a minimum effect size in the sample—while sensitiveness targets a minimum effect size in the sample without hoping to know the real effect size in the population.

Table 2

*Sample Sizes Required for Capturing Cohen's Small, Medium and Large Effects*

| test | ES | small | | medium | | large | |
|:---:|:---:|:---:|:---:|:---:|:---:|:---:|:---:|
| | | sns | pwr | sns | pwr | sns | pwr |
| t | r | **272** | 614 | **32** | 64 | **12** | 21 |
| t | d | **275** | 620 | **48** | 102 | **21** | 42 |
| $\chi^2(1)$ | $w_{(1df)}$ | **384** | 785 | **43** | 88 | **16** | 32 |
| $\chi^2(2)$ | $w_{(2df)}$ | **594** | 964 | **67** | 108 | **24** | 39 |
| $\chi^2(3)$ | $w_{(3df)}$ | **774** | 1091 | **87** | 122 | **32** | 44 |
| $\chi^2(4)$ | $w_{(4df)}$ | **947** | 1194 | **106** | 133 | **38** | 48 |
| $\chi^2(5)$ | $w_{(5df)}$ | **1091** | 1283 | **124** | 143 | **45** | 52 |
| F(1,dfd) | $f_{(2g)}$ | **389** | 788 | **66** | 128 | **29** | 52 |
| F(2,dfd) | $f_{(3g)}$ | **605** | 969 | **102** | 159 | **44** | 66 |
| F(3,dfd) | $f_{(4g)}$ | **789** | 1096 | **133** | 180 | **57** | 76 |
| F(4,dfd) | $f_{(5g)}$ | **957** | 1200 | **161** | 200 | **68** | 80 |
| F(5,dfd) | $f_{(6g)}$ | **1116** | 1290 | **188** | 216 | **80** | 90 |

*Note.* Sns (sensitiveness) = N needed for rejecting Fisher's $H_0$ for *sig* = 5%. *Pwr* (power) = N needed for rejecting Neyman-Pearson's $H_M$ for $\alpha$ = 5%, $\beta$ = 20%. *t* point biserial *r*, *t* for independent groups; $\chi^2$ for goodness-of-fit, *F* oneway ANOVA. All tests are one-tailed.

**A priori sensitiveness analyses**

When a researcher is limited to given sample sizes, a priori sensitiveness analyses provide information about the expected minimum effect sizes for significance, so that the researcher may ascertain whether his research goals are achievable or not.

- The first step is to calculate the critical value of a test that corresponds to the desired level of significance and sample size. This step can be conveniently performed with online calculators (e.g. DanielSoper.com).
- The second step is to calculate the effect size associated with above critical value by using the appropriate formulas (Supplemental Table 1).

Let's say we are interested in testing mean differences between independent groups using a conventional level of significance (*sig* = .05), a one-tailed test, and a sample size of *N* = 30. We can easily find out the corresponding critical value in a *t* distribution—*t*(28) = 1.70, *p* = .05—, then derive the minimum effect size that such sample size would be able to capture under the circumstances—Cohen's *d* = 0.64. We are then in a good position to query whether our research, intervention or programme is capable of delivering at least such medium-to-large effect size or, otherwise, whether we would be better off increasing our sample size in order to capture a smaller effect size. (Incidentally, this is the *MES* that the N = 30 rule-of-thumb was able to capture in the simulation.)

**Post-hoc sensitiveness analyses**

Post-hoc analyses can also be performed for sensitiveness. The formula is the following:

$$\% \text{ sensitiveness} = 100 \left( \frac{N_{actual}}{N_{min}} - 1 \right)$$

Positive values indicate over-sensitiveness and negative values indicate under-sensitiveness, both as a percentage of the ratio between actual and minimum sample size. As it happens with power, over-sensitiveness decreases the stipulated *MES*, something which may fool a researcher into accepting a smaller *MES* simply because the test turned out statistically significant. Still, it is less of a concern than under-sensitiveness, which can be used for concluding nothing about a test result precisely because it lacks enough sensitiveness for capturing the expected minimum effect size.

For example, the simulation described earlier was set up with enough sensitiveness (*N* = 43 per condition and study) to capture a medium-sized difference in frequencies between pairs of conditions (Cohen's *w* = .3). Therefore, we can expect the N = 30 rule-of-thumb to be under-sensitive, something which it turned out to be, as it recommended a sample size 37.5% smaller than that required for sensitiveness. On the other hand, power-based sampling can be expected to be over-sensitive, which also turned out to be, as it recommended a sample size 112.5% larger than that needed for sensitiveness.

**Discussion and conclusions**

By now, the reader has probably put the data presented in Table 2 to test using G*Power and found that the sample size required for sensitiveness equals about 50% power (e.g., $N = 48$ for a *t* test and medium-sized *d* equals 52% power, and $N = 43$ for a $\chi^2$ test with one degree of freedom and medium-sized *w* equals 50% power). Does this mean that the sensitiveness analysis proposed in this article ultimately fails the research community by encouraging underpowered research? The answer to this riddle is 'No', for several reasons now discussed.

Firstly, as illustrated in Figure 1, sensitiveness and power are related as they both share location with the critical value of the test. Indeed, our reader can derive the effect size of the critical *t* used for a power-set test (e.g., for a medium-sized *d* and 80% power) and realize that it does not result in the 'expected' $d = 0.5$ but $d = 0.37$ (the *MES*). This is so because Cohen's post hoc formulas (see Supplemental Table 1) attempt to estimate unknown population effect sizes—that is, the middle of the unknown alternative distribution—which tend to comprise about 50% power. Thus, the lower 'power' of sensitiveness is due to such formulation goal rather than to an inherent lack of power. In any case, if the reader goes down the table and onto more complex tests (i.e., those with more groups to take into account), he will find that as the number of groups increases so does the power. That is, the power equivalent to a sensitive medium-sized *w* with three degrees-of-freedom—$\chi^2(3)$—is 64%, while the power equivalent to a sensitive medium-sized *f* with six groups—$F(5,dfd)$—is 74%. Again, sensitiveness does not always imply low power.

Secondly, although sensitiveness and power are related, they are not necessarily the same. Indeed, we could define power in terms of sensitiveness—80% power implies that the research sample is large enough for a test to be sensitive to 80% of effect sizes, i.e. those above *MES*—but not vice versa—a test may be sensitive and turn out significant despite lacking power (e.g., Button et al., 2013; Maxwell, Lau, & Howard, 2015; Vadillo, Konstantinidis, & Shanks, 2015).

Thirdly, power analysis assumes a known effect size, provided by the alternative hypothesis. Thus, any time such effect size is either known or can be reasonably set with some accuracy, power is the best sampling strategy to use, precisely because it allows for good control of the research context. Sensitiveness, on the other hand, is most appropriate when the researcher does not know the effect size in the population or when he is after a particular minimum effect size. This seems to be the case with most current research, which tend to be pilot studies, pioneer research or overly based on Fisher's approach (including most research done under the banner of null hypothesis significance testing). This is to say, if research is oriented towards achieving a desired minimum effect size, including in the vague form of a statistical result beyond a threshold of significance which may be used as evidence against a single null hypothesis, then sensitiveness is the most appropriate tool to use for sample estimation.

Lastly, a larger sample will always trump a smaller one. If a researcher can afford a sample size in the thousands, she may as well discard both power and sensitiveness as a matter of concern (another issue, of course, is the practical significance of the results thereby obtained). However, if the researcher wishes to tailor the sample size to either a known effect size in the population or a desired minimum effect size in the sample, then power analysis and sensitiveness analysis, respectively, will be the appropriate strategies to use.

In summary, I see sensitiveness as an important advance in the mathematical estimation of sample sizes and fitter-for-purpose than power analyses when using Fisher's tests of significance. Thus, researchers following Fisher's approach will benefit from estimating sample size for sensitiveness instead of for power.

Furthermore, what I present here is merely a statistical technology. The philosophical differences between Neyman-Pearson's and Fisher's approaches are important and need to be considered for understanding the construct of effect size in the context of sensitiveness— namely as the effect size expected in the population of samples similar to the research sample (Fisher, 1955). Because philosophical considerations are often ignored in research work (Hager, 2013), I decided not to emphasise them in this article.

STATISTICAL SENSITIVENESS - page 17

## Supplemental material 1

Supplemental Table 1

*Effect Size Formulas and Cohen's Conventional Effect Size Values*

| | |
|---|---|
| $r = r$ | $r$ ; small = 0.10; medium = 0.30; large = 0.50 |
| $d = \dfrac{2t}{\sqrt{df}}$ | $d$ ; small = 0.20; medium = 0.50; large = 0.80 |
| $w = \sqrt{dfs}\sqrt{\chi^2 \div (dfs\ N)}$ | $w$ ; small = 0.10; medium = 0.30; large = 0.50 |
| $f = \dfrac{2\sqrt{(dfn\ F) \div dfd}}{2}$ | $f$ ; small = 0.10; medium = 0.25; large = 0.40 |

## Supplemental material 2

Supplemental Table 2

*Target and Actual Effect Sizes, Sample Sizes and Critical Test Values*

| target ES | | N | CV (sig < .05) | actual ES |
|---|---|---|---|---|
| small | r = .1 | 272 | t(270) = 1.6514 | r = .1000 |
| medium | r = .3 | 32 | t(30) = 1.7225 | r = .3000 |
| large | r = .5 | 12 | t(10) = 1.8257 | r = .5000 |
| small | d = 0.2 | 275 | t(273) = 1.6505 | d = 0.1998 |
| medium | d = 0.5 | 48 | t(46) = 1.6787 | d = 0.4950 |
| large | d = 0.8 | 21 | t(19) = 1.7291 | d = 0.7934 |
| small | w = φ = .1 | 384 | $\chi^2(1) = 3.8415$ | φ = .1000 |
| medium | w = φ = .3 | 43 | $\chi^2(1) = 3.8415$ | φ = .2989 |
| large | w = φ = .5 | 16 | $\chi^2(1) = 3.8415$ | φ = .4900 |
| small | w = V(2) = .071 | 594 | $\chi^2(2) = 5.9915$ | V(2) = .0710 |
| medium | w = V(2) = .212 | 67 | $\chi^2(2) = 5.9915$ | V(2) = .2115 |
| large | w = V(2) = .354 | 24 | $\chi^2(2) = 5.9915$ | V(2) = .3533 |
| small | w = V(3) = .058 | 774 | $\chi^2(3) = 7.8147$ | V(3) = .0580 |
| medium | w = V(3) = .173 | 87 | $\chi^2(3) = 7.8147$ | V(3) = .1730 |
| large | w = V(3) = .289 | 32 | $\chi^2(3) = 7.8147$ | V(3) = .2853 |
| small | w = V(4) = .050 | 947 | $\chi^2(4) = 9.4877$ | V(4) = .0500 |
| medium | w = V(4) = .150 | 106 | $\chi^2(4) = 9.4877$ | V(4) = .1496 |
| large | w = V(4) = .250 | 38 | $\chi^2(4) = 9.4877$ | V(4) = .2498 |
| small | w = V(5) = .045 | 1091 | $\chi^2(5) = 11.0705$ | V(5) = .0450 |
| medium | w = V(5) = .134 | 124 | $\chi^2(5) = 11.0705$ | V(5) = .1336 |
| large | w = V(5) = .224 | 45 | $\chi^2(5) = 11.0705$ | V(5) = .2218 |
| small | f = 0.10 | 389 | F(1,387) = 3.8656 | f = 0.0999 |
| medium | f = 0.25 | 66 | F(1,64) = 3.9909 | f = 0.2497 |
| large | f = 0.40 | 29 | F(1,27) = 4.2100 | f = 0.3949 |
| small | f = 0.10 | 605 | F(2,602) = 3.0107 | f = 0.1000 |
| medium | f = 0.25 | 102 | F(2,99) = 3.0882 | f = 0.2498 |
| large | f = 0.40 | 44 | F(2,41) = 3.2257 | f = 0.3967 |

| target ES | | N | CV (sig < .05) | actual ES |
|---|---|---|---|---|
| small | f = 0.10 | 789 | F(3,785) = 2.6162 | f = 0.1000 |
| medium | f = 0.25 | 133 | F(3,129) = 2.6748 | f = 0.2494 |
| large | f = 0.40 | 57 | F(3,53) = 2.7791 | f = 0.3966 |
| | | | | |
| small | f = 0.10 | 957 | F(4,952) = 2.3813 | f = 0.1000 |
| medium | f = 0.25 | 161 | F(4,156) = 2.4296 | f = 0.2496 |
| large | f = 0.40 | 68 | F(4,63) = 2.5177 | f = 0.3998 |
| | | | | |
| small | f = 0.10 | 1116 | F(5,1110) = 2.2222 | f = 0.1000 |
| medium | f = 0.25 | 188 | F(5,182) = 2.2638 | f = 0.2494 |
| large | f = 0.40 | 80 | F(5,74) = 2.3383 | f = 0.3975 |

**Supplemental material 3**

**Simulation environment**
The overall simulation strategy aimed to recreate a typical research context with the following characteristics:

- Single research projects rather than repeated sampling from the same population. The overall research thus comprised a total of 1,032 tests carried out on 1,032 independent samples extracted from 344 independent populations of differing size. The populations and samples were simulated as follows:
    - In order to prevent repeated sampling from the same population, four macro-populations were generated for the simulation. These macro-populations run from large to small ($N = 20,000$, $N = 10,000$, $N = 4,000$, and $N = 2,000$), comprised two equal-sized independent groups, and were randomly generated in Excel (with group parameters $N$~10,1 and $N$~10.5,1) as for representing a medium-effect size difference between groups (Cohen's $d = 0.5$). (We may think about macro-populations as representing psychological disorders with different prevalence internationally (e.g., major depression, rather prevalent, and schizophrenia, rather rare). Because they are international macro-populations, they are inaccessible (research is done on local populations, instead); and because they are inaccessible, the real effect sizes of disorders, although assumed fixed, are unknown. This is why simulated macro-populations with known parameters are not suitable for testing Fisher's approach as such, thus the next step.)

    - In order to increase local uncertainty regarding the parameters of the macro-populations, independent research populations were randomly extracted from above macro-populations using SPSS. The extraction criteria were twofold: size ($N = 2,000$ and $N = 1,000$ were extracted from macro-population 20,000; $N = 2,000$ and $N = 500$ from macro-population 10,000; $N = 1,000$ and $N = 200$ from macro-population 4,000; and $N = 500$ and $N = 200$ were extracted from macro-population 2,000) and independence (each research population was created anew for each round, totalling 43 independent populations per study and 344 populations overall). (We may think about populations as representing prevalence of psychological disorders at regional levels. Because it cannot be reasonably expected for all disorders to be proportionally distributed around the world, but, perhaps, that personality disorders are more prevalent in one country while schizophrenia is more prevalent in another, the populations were not necessarily extracted in a manner proportional to the size of the macro-populations. Furthermore, it cannot be expected that researchers from different parts of a region will access the entire population but, instead, that they will access local subsets of such population, even when the intention is to generalize to the whole population. In order to mimic this, each population was extracted independently, further increasing the uncertainty about the parameters for each local population. Notice that at this point, the exact effect size for each local population as well as the exact means and standard deviations per group are unknown a priori, which is what the exercise

attempted to achieve—i.e., although it is reasonable to think that they are not far from the initial known parameters, it is impossible to say with certainty. Indeed, Supplemental Table 3 provides summary statistics for populations at "regional" levels, starting to show random deviations from the initial parameters; consequently, greater uncertainty is expected for each individual "local" population, from which the samples were taken.)

- o To better mimic the single-study ad hoc approach of a typical research project, research samples were then randomly selected from each population per study condition ($k = 3$) using SPSS, totalling 1,032 samples overall.

Supplemental Table 3

*Descriptive Statistics for Eight Simulated Populations*

| Statistic | I | II | III | IV | V | VI | VII | VIII |
|---|---|---|---|---|---|---|---|---|
| N 1 | 5000 | 10000 | 2000 | 2000 | 1000 | 10000 | 5000 | 1000 |
| Mean 1 | 10.00 | 9.99 | 10.01 | 9.97 | 10.03 | 10.00 | 10.00 | 10.03 |
| SD 1 | 1.00 | 1.00 | 0.98 | 1.02 | 1.02 | 1.00 | 1.00 | 1.02 |
| N 2 | 5000 | 10000 | 2000 | 2000 | 1000 | 10000 | 5000 | 1000 |
| Mean 2 | 10.48 | 10.51 | 10.49 | 10.52 | 10.48 | 10.48 | 10.48 | 10.48 |
| SD 2 | 1.01 | 0.99 | 1.01 | 1.01 | 1.02 | 0.99 | 1.01 | 1.02 |
| d | 0.48 | 0.53 | 0.48 | 0.55 | 0.45 | 0.49 | 0.48 | 0.45 |
| Research N | 2000 | 2000 | 1000 | 200 | 200 | 1000 | 500 | 500 |
| PWR, % sig | 72.1 | 86.0 | 79.1 | 90.7 | 69.8 | 76.7 | 86.0 | 74.4 |
| SNS, % sig | 55.8 | 48.8 | 37.2 | 53.5 | 44.2 | 58.1 | 53.5 | 37.2 |
| THMB, % sig | 23.3 | 37.2 | 37.2 | 44.2 | 30.2 | 44.2 | 41.9 | 32.6 |

*Note*. N = Macro-population size (per group), Research N = total size of each of 43 independent research populations (each generated in order to draw a single random sample per condition). % sign = percentage of significant tests irrespective of effect size considerations, doubling as an indicator of power. PWR = power-based sampling ($n = 102$), SNS = sensitiveness-based sampling ($n = 48$), THMB = rule-of-thumb-based sampling ($n = 30$).

- The simulation also tried to mimic research contexts when, despite the real effect size in the population being unknown, a researcher nonetheless decides to use an effect size both for purposes of sampling estimation and of result interpretation. The minimum effect size considered meaningful, reasonable or defendable was set, "coincidentally", to be Cohen's $d = 0.5$, subject to the statistical significance of a test. (The minimum effect size ought to be the same than the effect size designed for the simulation at macro-population levels in order to make a meaningful comparison between conditions without unnecessary extraneous complications.) Larger effect sizes which were also statistically significant were, or course, acceptable, thus counted, while smaller effect sizes were considered untenable and, thus, ignored irrespectively of their statistical significance. (This procedure gives a greater role to practical significance than to statistical significance; otherwise, a larger sample will always be more powerful and, thus, return a significant result for smaller effects, even when the researcher had decided a priori that those effects were unimportant. That is, while a larger sample always trumps a smaller one, statistical significance typically trumps effect sizes, which is what the research tried to avoid.)

- The conditions under test were the three sampling strategies discussed in the main article:
  - Power-based sampling. Conventional parameters were plugged into G*Power: one-tailed *t* test (between groups), *ES* = 0.5, *α* = .05, power = .80, ratio allocation = 1, which recommended a total sample of *n* = 102 (i.e., *n* = 51 per group). This condition thus represents a reasonable procedure followed by a researcher who does not know, and does not wish to double-guess, the real effect size in the population, yet is still interested in a particular minimum effect size. Because Cohen (1988) does not talk about *MES*, such researcher only has the resource of plugging in the desired effect size and of later ignoring a significant result whose effect size is smaller than desired.

  - Sensitiveness-based sampling. The a priori procedure for sensitiveness was run also using conventional parameters: *t* test, *sig* = .05, and *ES* = 0.5. The minimum total sample size recommended was *n* = 48 (i.e., *n* = 24 per group). This condition thus represents the sampling approach discussed in this document: carrying out a significance test but ensuring that it is sensitive enough to capture the desired (minimum) effect size. Because this desired *MES* is independent of the unknown population effect size, a researcher may carry on without concerning herself with double-guessing the later.

  - Rule-of-thumb sampling. Using *n* = 30, this condition represents the use of a rule-of-thumb independently of the particularities of the research project. In aiming for sound yet economic sampling, a researcher may opt for this rule-of-thumb, as it "promises" normal sampling distributions adequate for *t*-tests (e.g., Crawley, 2014).

- Analyses focused on the frequency of significant *t* tests which captured the specified effect size or larger. The difference in counts between pairs of conditions were tested using $\chi^2$ for goodness-of-fit. Therefore, the effect size took precedence over statistical significance so that a *t* test was not counted if the effect size was smaller than expected (i.e., *d* < 0.495, for rounding purposes) or if the test was not significant. The target *MES* for $\chi^2$ was a medium effect size (*w* = .30), which recommended a minimum sample size of *N* = 43 per condition and study.

The overall simulation was broken down into eight studies, each study including 43 independent research populations of similar size and comprising 129 independent samples. The overall simulation amounted to 1,032 samples. Supplemental Table 3 collates the main descriptive statistics for the eight studies.

Supplemental Table 4 (as well as Table 1) collates the main results of the simulation (frequencies, percentages, $\chi^2$ tests, and effect sizes). As can be observed, most paired comparisons turned out to be below the set threshold for a minimum effect of moderate size ($w = .30$), and the only one which turned out statistically significant was just about moderate in size ($w = .33$). The aggregated total further illustrates that the difference between power and sensitiveness-based sampling is rather negligible ($w = .03$) when a researcher seeks a minimum effect size and not merely statistical significance.

Surprisingly enough, the N = 30 rule-of-thumb did not fare too bad either. Although it certainly shows a small effect size ($w \geq .10$), it fared relatively well under the criterium used for the simulation in the first place. In any case, given the random selection of samples, we can be confident that no bias other than random variability was present in the simulation, and we may deem the effect sizes obtained reliable despite the lack of statistical significance. Thus, we can conclude that the sampling strategies using power and sensitiveness were sensibly more beneficial to the research than the N = 30 rule-of-thumb in the case of this simulation.

Supplemental Table 4

*Simulation Results (Frequencies, Percentages, Tests, and Effect Sizes)*

| Statistic | I | II | III | IV | V | VI | VII | VIII | All studies |
|---|---|---|---|---|---|---|---|---|---|
| f PWR | 17 | 18 | 21 | 28 | 16 | 19 | 23 | 16 | 158 |
| f SNS | 20 | 21 | 13 | 22 | 18 | 20 | 20 | 15 | 149 |
| f THMB | 10 | 16 | 16 | 18 | 13 | 18 | 17 | 14 | 122 |
| % PWR | 36.2 | 32.7 | 42.0 | 41.2 | 34.0 | 33.3 | 38.3 | 35.6 | 36.8 |
| % SNS | 42.6 | 38.2 | 26.0 | 32.4 | 38.3 | 35.1 | 33.3 | 33.3 | 34.7 |
| % THMB | 21.3 | 29.1 | 32.0 | 26.5 | 27.7 | 31.6 | 28.3 | 31.1 | 28.4 |
| $\chi^2$ PWR-SNS | 0.24 | 0.23 | 1.88 | 0.72 | 0.12 | 0.03 | 0.21 | 0.03 | 0.26 |
| w | 0.08 | 0.08 | 0.24 | 0.12 | 0.06 | 0.03 | 0.07 | 0.03 | 0.03 |
| $\chi^2$ SNS-THMB | 3.33 | 0.68 | 0.31 | 0.40 | 0.81 | 0.11 | 0.24 | 0.03 | 2.69 |
| w | 0.33 | 0.14 | 0.10 | 0.10 | 0.16 | 0.05 | 0.08 | 0.03 | 0.10 |
| $\chi^2$ THMB-PWR | 1.82 | 0.12 | 0.68 | 2.17 | 0.58 | 0.87 | 0.34 | 0.72 | 4.63 |
| w | 0.26 | 0.06 | 0.14 | 0.22 | 0.14 | 0.15 | 0.09 | 0.15 | 0.13 |

*Note.* Statistics for significant tests capturing d > 0.495. Each study tested 43 independent samples per condition (i.e., 129 samples per study, and 1,032 samples in total). *f* = frequencies for significant tests capturing d > 0.495; PWR = power-based sampling ($n = 102$), SNS = sensitiveness-based sampling ($n = 48$), THMB = rule-of-thumb-based sampling ($n = 30$). *P*-values intentionally omitted, as they offer less valuable information than target minimum effect size ($w = .30$).